\documentclass[journal,draftclsnofoot,onecolumn,12pt]{IEEEtran}

\usepackage{acronym}
\usepackage{amsfonts}
\usepackage[dvips]{graphicx}
\usepackage{times}
\usepackage{cite}
\usepackage{amsmath}
\usepackage{array}
\usepackage{amssymb}
\usepackage{stfloats}
\usepackage{diagbox}
\usepackage{graphicx}
\usepackage{footnote}
\usepackage{amsthm}
\usepackage{booktabs}
\usepackage{array}
\usepackage[ruled,vlined]{algorithm2e}
\usepackage{subeqnarray}
\usepackage{cases}
\usepackage{threeparttable}
\usepackage{color}
\usepackage{epstopdf}
\usepackage{multirow}
\usepackage{tabularx}
\usepackage{enumerate}
\usepackage{multicol}
\usepackage{hyperref}
\usepackage{subfig}
\usepackage{caption}

\def\bb0{{\mathbb{0}}}

\def\bb{{\mathbf{b}}}

\def\b0{{\mathbf{0}}}

\def\sf0{{\mathsf{0}}}

\def\rm0{{\mathrm{0}}}

\acrodef{CSI}[CSI]{channel state information}
\acrodef{CSIT}[CSIT]{channel state information at the transmitter}
\acrodef{CSIR}[CSIR]{channel state information at the receiver}
\acrodef{MIMO}[MIMO]{multiple-input multiple-output}
\acrodef{SISO}[SISO]{single-input single-output}
\acrodef{MISO}[MISO]{multiple-input single-output}
\acrodef{SIMO}[SIMO]{single-input multiple-output}
\acrodef{ADCs}[ADCs]{analog-to-digital convertors}
\acrodef{SNR}[SNR]{signal-to-noise ratio}
\acrodef{AWGN}[AWGN]{additive white Gaussian noise}
\acrodef{MRT}[MRT]{maximal ratio transmission}
\acrodef{DFT}[DFT]{Discrete Fourier Transform}
\acrodef{ULA}[ULA]{uniform linear array}
\acrodef{UPA}[UPA]{uniform planar array}
\acrodef{LS}[LS]{least squares}
\acrodef{ALMMSE}[ALMMSE]{approximate linear minimum mean squared error}
\acrodef{QIHT}[QIHT]{quantized iterative hard thresholding}
\acrodef{QIST}[QIST]{quantized iterative soft thresholding}
\acrodef{SVD}[SVD]{singular value decomposition}

\ifCLASSINFOpdf

\else
\fi
\begin{document}

\title{Model-Driven Deep Learning for Physical Layer Communications}

\author{Hengtao He, Shi Jin, Chao-Kai Wen, Feifei~Gao, \\
Geoffrey Ye Li, and Zongben Xu

\thanks{H.~He and S.~Jin are with the National Mobile Communications Research
Laboratory, Southeast University, Nanjing 210096, China (e-mail: hehengtao@seu.edu.cn, and jinshi@seu.edu.cn).}
\thanks{C.-K.~Wen is with the Institute of Communications Engineering, National
Sun Yat-sen University, Kaohsiung 804, Taiwan (e-mail: chaokai.wen@mail.nsysu.edu.tw).}
\thanks{F. Gao is with Institute for Artificial Intelligence, Tsinghua University (THUAI);
State Key Lab of Intelligent Technologies and Systems, Tsinghua University;
Beijing National Research Center for Information Science and Technology (BNRist);
Department of Automation, Tsinghua University,
Beijing, P.R. China (email: feifeigao@ieee.org).}

\thanks{G.~Y.~Li is with the School of Electrical and Computer Engineering,
Georgia Institute of Technology, Atlanta, GA 30332 USA (e-mail:
liye@ece.gatech.edu).}

\thanks{
Z. Xu is with the Institute for Information and
System Sciences, Xian Jiaotong University, Xi'an 710049, China (e-mail:
zbxu@mail.xjtu.edu.cn).}

}

\maketitle
\vspace{-1cm}
\begin{abstract}

Intelligent communication is gradually  becoming a  mainstream direction. As a major branch of machine learning, deep learning (DL) has been applied in physical layer communications and demonstrated an impressive performance improvement in recent years. However, most of the existing works related to DL focus on data-driven approaches, which consider the communication system as a black box and train it by using a huge volume of data. Training a network requires sufficient computing resources and extensive time, both of which are rarely found in communication devices. By contrast, model-driven DL approaches combine communication domain knowledge with DL to reduce the demand for computing resources and training time. This article discusses the recent advancements in  model-driven DL approaches in physical layer communications, including transmission schemes, receiver design, and channel information recovery. Several open issues for future research are also highlighted.

\end{abstract}


%
\IEEEpeerreviewmaketitle

\section{Introduction}
Modern wireless communication systems have developed from the first to the forth generation and have propelled to the fifth generation ($5$G) to provide advanced wireless services, such as virtual reality, autonomous driving, and Internet of Things (IoT). Enhanced mobile broadband, massive machine-type communications, and ultra-reliable and low-latency communications are the three main scenarios for $5$G wireless networks. They require communication systems to have the ability to handle a large amount of wireless data, recognize and dynamically adapt to complex environments, and satisfy the requirements for high speed and accurate processing. Therefore,  future wireless communication systems  must be highly intelligent.

As a prevailing approach to artificial intelligence, machine learning, specifically deep learning (DL), has drawn much attention in recent years because of its great successes in computer vision, natural language processing. Recently, it has been applied to wireless communications, such as physical layer communications\cite{DL2017wang,DL2018Qin,DL2018Gui,DL18DOA}, resource allocation\cite{DL19Ye,DL2018SDN}, and intelligent traffic control\cite{DL2018traffic}. However, most of the existing DL networks are  data-driven, which use the standard neural network structure as a black box and train it by a large amount of data.
Training a standard neural network also requires a long training time in addition to a huge data set, which is often scarce, especially in wireless communications. In contrast to  fully data-driven methods, model-driven DL approaches construct the network topology based on known physical mechanism and domain knowledge, and therefore require less training data and shorter training time \cite{modelDL}. They have begun to apply in physical layer communications and become promising for achieving intelligent communications.


This article elaborates the viewpoint of model-driven DL, provides a comprehensive overview on its applications in physical layer communications, and highlights promising areas for future research. The rest of this article is organized as follows. Section \ref{Model} introduces the basic principles of model-driven DL in physical layer communications. Then, Sections \ref{Transmission} and \ref{receiver} present the model-driven-DL-based transmission schemes and receiver design, respectively. Section \ref{CSI} demonstrates data-driven DL methods are sometimes indispensable to address inaccurate models, such as channel state information (CSI) recovery. Section \ref{Open} discusses some open issues in this research area before the article is concluded in Section \ref{con}.

\section{Why model-driven DL ?}\label{Model}
Using a fully data-driven DL approach is often the most popular solution in the field of computer vision and natural language processing  since the mathematical description of the task cannot be easily  obtained in these fields. Without relying on a mathematical model and expert knowledge, a data-driven neural network can be designed by using standard, usually fully-connected neural network and the hyperparameters can be tuned by conducting engineering experiments. However, the performance of the data-driven approaches heavily depends on a huge amount of labeled data, as the network cannot learn much insight if the training set is small. However, labeled data cannot be easily obtained in some applications, especially in wireless communications. Furthermore, lack of a theoretical understanding about the relationship between neural network topology and performance makes its structure unexplainable and unpredictable. These limitations prevent the widespread use of data-driven DL approaches in some practical applications.

To address this issues, model-driven DL has been proposed to make a network explainable and predictable \cite{modelDL}.  The main characteristics of model-driven DL are the network constructed based on domain knowledge, rather than heavily depends on the huge volume of labeled data to choose suitable standard neural network. The domain knowledge in physical layer communications is in the form of the models and algorithms developed over several decades of intense research.

\begin{figure}[h]
	\centering
	\includegraphics[width=6in]{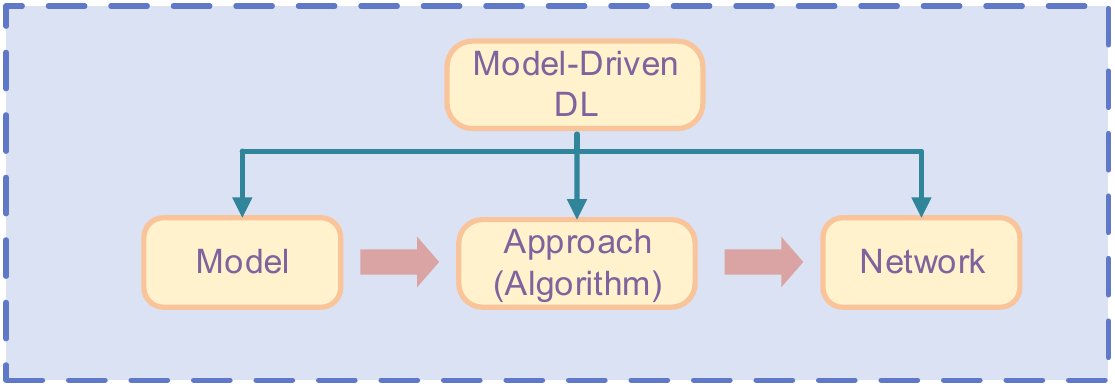}
	\caption{.~~Components of the model-driven DL approach.}
	\label{model_driven}
\end{figure}

As illustrated in Fig.\,\ref{model_driven}, the model-driven DL comprises three parts: a model, an approach (algorithm), and a network for a specific task. The model is constructed based on the background knowledge, including the physical mechanisms and domain knowledge. Different from the model in the analytical approach, the one in model-driven DL only provides a very rough and broad definition of the solution, thereby reducing the pressure for accurate modelling. An approach (i.e., feasible algorithm and theory) based on the aforementioned model and domain knowledge is then designed to solve the problem. This approach is then employed to obtain a deep network with several learnable parameters that are trained via back-propagation  algorithm similar to data-driven DL. In many cases, the network is constructed by unfolding an iterative algorithm into a signal flow graph, using the accessible algorithm as an initialization step and combining it with a standard neural network, or mimicking the conventional structure of the model-driven method.

Model-driven DL approaches are attractive for physical layer communications as the mathematical models are often available even though possibly simplified and inaccurate. The domain knowledge  acquired over several decades of intense research in wireless communications can still be exploited. Model-driven DL inherits the advantages of the model-driven approaches and avoids the requirements for accurate modeling. The imperfections resulting from the inaccuracy of the model and predetermined parameters can  be compensated by the powerful learning ability of DL. Furthermore, the model-driven DL has other benefits, such as its low demand for training data, reduced risk of overfitting, and rapid implementation. Therefore, model-driven DL 
is a promising method for intelligent communications. This article will explore several aspects of model-driven DL for physical layer communications in detail.

\section{Transmission Schemes}\label{Transmission}
An autoencoder in DL usually maps the original data from the input layer into a code, which can recover the data that closely matches the original one at the output layer. By mimicking such a concept, the method in \cite{OShea} considers the communication transceiver design as an autoencoder task. As illustrated in Fig.\,2, the transmitter and receiver are represented by fully connected deep neural networks (DNN) and are jointly optimized over an AWGN channel. The end-to-end learning communication system based on the autoencoder is purely data-driven, which treats the communication system as a black box. The purely data-driven method can potentially do better than the conventional handmade design. The autoencoder-based system has been extended to multi-user communication over interference channels, orthogonal frequency division multiplexing (OFDM) systems with multipath channels \cite{OFDMautoencoder}. In \cite{GAN}, the purely data-driven DL method has been used in transceiver optimization without using accurate CSI. To avoid the black box architecture, expert knowledge has been incorporated into autoencoder-based communication systems, with the radio transformer network (RTN) \cite{OShea} and the OFDM-autoencoder \cite{OFDMautoencoder} as two representative examples.

\subsection{RTN}\label{end2end}
The RTN incorporates communication expert knowledge into the network and can augment signal processing ability and accelerate the training phase \cite{OShea}. It can be regarded as a simple model-driven DL network that comprises a learned parameter estimator, a parametric transform, and a learned discriminative network. By adding a parameter estimation module to the original autoencoder-based communication system in the receiver, the RTN architecture can adapt to complex scenarios with hardware impairments. As demonstrated in \cite{OShea}, the autoencoder with the RTN consistently outperforms and converges faster than the original autoencoder, thereby highlighting the power of expert knowledge in DL.

\subsection{Model-Driven DL for OFDM Transmission}\label{end2end}
As a standard transmission technology, the cyclic-prefix (CP)-based OFDM scheme can combat multipath fading. In \cite{OFDMautoencoder}, the structure of the autoencoder has combined with the OFDM system and the end-to-end approach has been used to train the whole network. Compared with completely plain autoencoder-based communication systems \cite{OShea}, the OFDM-autoencoder in \cite{OFDMautoencoder}  extends the flexible structure of the autoencoder into a conventional OFDM system. This neural network inherits the advantages of OFDM systems, such as  robust against sampling synchronization errors, and simplifying the process of equalization over multipath channels. The OFDM-autoencoder combined with the RTN can further handle carrier frequency offset directly in the time domain with a slight performance degradation while the hardware impairments can be easily compensated by carefully design.

To address high peak-to-average power ratio (PAPR) in OFDM systems, a PAPR reducing network (PRNet) has been proposed in \cite{OFDMPAPR}. This scheme adaptively learns the optimal constellation mapping and demapping via model-driven DL. Specifically, the PRNet embeds the encoder and decoder to a conventional structure of the OFDM system but still reserves several traditional blocks, such as the fast Fourier transform (FFT) module. To reduce PAPR and prevent the deterioration in bit-error rate (BER), the training is divided into two stages with these two distinct objectives. From \cite{OFDMPAPR}, the PRNet outperforms conventional schemes in terms of PAPR and BER.

\section{Receiver Design}\label{receiver}
In recent years, DL has been applied to receiver design because of its power in signal processing. By taking the OFDM receiver and MIMO detection as examples, we will demonstrate how model-driven DL can be applied in receiver design in this section.
\subsection{Model-Driven DL for OFDM Receiver Design}\label{OFDM_sec}
OFDM has been widely used in various communication systems to combat frequency-selective fading and to achieve multi-Gbps transmission. In a traditional OFDM system, the signal is processed in transceivers in a block-by-block manner. Specifically, the receiver utilizes the known pilot and the received symbols for channel estimation (CE) and signal detection (SD), respectively.

With the emergence of the widely adapted DL, a new OFDM receiver architecture has been introduced in \cite{8052521}. This architecture adopts a five-layer fully connected DNN (FC-DNN) and replaces all blocks at the traditional OFDM receiver, including FFT, CE, SD, and QAM demodulation. Such a data-driven approach treats the receiver as a black box, exploits no expert knowledge in wireless communications, and results in the FC-DNN-based receiver unexplainable and unpredictable. As in all data-driven methods, it depends on a huge amount of data to train a large number of parameters. Therefore, the architecture converges slowly and shows a relatively high computational complexity.


Model-driven DL can be integrated into the OFDM receiver design to address the above issues. Since all blocks in OFDM transceivers have been rigorously developed, the existing algorithms can be configured as the fundamentals of the model family in model-driven DL. In \cite{ComNet}, a model-driven DL based OFDM receiver, called ComNet, has been proposed, which combines DL with expert knowledge. As shown in Fig.\,3, the ComNet receiver involves two subnets for CE and SD and is constructed in a block-by-block manner similar to the conventional communication systems. The initialization of each subnet is an accessible communication algorithm.

This model-driven DL approach provides more accurate CE compared with linear minimum mean-squared error (LMMSE) CE under linear and nonlinear cases (e.g., CP removal and clipping) and outperforms both the traditional MMSE method and the FC-DNN in \cite{8052521}. In terms of complexity, this model-driven DL approach converges relatively faster and requires fewer parameters than the FC-DNN OFDM receiver \cite{8052521}. This approach has also been proven to be robust to signal-to-noise ratio (SNR) mismatch. These advantages indicate that the expert knowledge in wireless communications  can be effectively used to boost the performance of the OFDM receiver with model-driven DL.
\vspace{-0.5cm}
\subsection{Model-Driven DL for MIMO Detection}\label{MIMO}

MIMO is a critical technique in  fourth generation ($4$G) cellular systems and  wireless area networks owing to its ability to increase both spectral efficiency and link reliability. The performance and complexity of a MIMO detection algorithm play important roles in receiver design. Given their superior performance and moderate complexity, iterative detectors, such as approximate-message-passing- (AMP) and expectation-propagation-based detectors, have attracted increasing attention. However, these iterative detectors are based on the assumption that channels follow a particular distribution and are hard to demonstrate an optimal performance under many complicated environments (e.g., Kronecker-correlated and Saleh-Valenzuela channels). Moreover, these detectors have fixed generic architecture and parameters while the incorporation of data knowledge remains unknown, thereby restricting the performance of iterative algorithms.

To address the above issues, a model-driven DL approach, called detection network (DetNet), has been designed in \cite{DetNet} by applying a projected gradient descent method for detecting maximum likelihood in a neural network. DetNet recovers the transmitted signal by treating the received signal and perfect CSI as inputs. This approach outperforms the iterative algorithms and is comparable with the K-best sphere decoder in terms of its symbol-error rate (SER) performance. The robustness of DetNet to some complex channels, such as the fixed channel with a deterministic ill condition and the varying channel model with a known distribution, are also demonstrated. DetNet obtains accuracy rates that are equal to and greater than those of the AMP and semidefinite relaxation algorithms, respectively.

To further reduce the number of the parameters that need to be trained, a model-driven DL network, called orthogonal AMP (OAMP)-Net, has been developed in \cite{OAMPNet} for MIMO detection via unfolding the iterative OAMP detector. Compared with the OAMP algorithm, OAMP-Net introduces only a few trainable parameters to incorporate the side information from the data. The number of trainable variables is independent of the number of antennas and is only determined by the number of layers, thereby giving OAMP-Net an advantage in addressing the large-scale problems, such as massive MIMO detection. With only few trainable variables, the stability and speed of convergence can be significantly improved in the training process. The network can also handle a time-varying channel with only a single training and improve the performance of the OAMP detector in Rayleigh and correlated MIMO channels.

\section{CSI Estimation and Feedback}\label{CSI}

In the previous two sections, we have discussed model-driven DL for transmission schemes and receiver design, respectively. Model-driven DL can be also used in CSI estiamtion and feedback, which is related to the whole transceiver design. For massive MIMO systems, achieving the potential advantages heavily depends on  accurate CSI. However, CSI estimation and feedback are challenging problems in massive MIMO systems as an accurate and specific channel model cannot be easily obtained. Under this circumstance, a large amount of data must be used to allow the neural network to capture the feature of massive MIMO channels. By taking channel estimation and feedback as examples, this section examines CSI estimation and feedback based on model-driven DL.

\subsection{Model-Driven DL for Beamspace Channel Estimation}

MmWave massive MIMO enables the use of multi-GHz bandwidth and large antenna arrays to offer high data rates and is thereby considered an important technique in future wireless communications. However, using a dedicated radio-frequency (RF) chain for each antenna may entail the use of expensive hardware or a very high power consumption. The lens-antenna-array-based beamspace mmWave system aims to reduce the number of RF chains and has been regarded as a promising architecture for future mmWave communications. Nevertheless, channel estimation for beamspace mmWave massive MIMO systems is extremely challenging, especially when there are a large antenna array and limited number of RF chains.

In \cite{CE}, the learned denoising-based AMP (LDAMP)-based channel estimation method  has been used in the beamspace mmWave massive MIMO system. Different from \cite{DL18DOA} that proposes a DNN-based super-resolution channel estimation method, the channel matrix is regarded as a $2$D natural image in \cite{CE} and a LDAMP neural network is developed for channel estimation, in which the denoising convolutional neural network is incorporated into the AMP algorithm. The LDAMP network is based on the compressed signal recovery model and the iterative signal recovery algorithm, which can be regarded as a model-driven DL network. The network learns the channel structure, estimates the channel from a large number of training data, and demonstrates an excellent performance even with a small number of RF chains at the receiver.  Furthermore, the performance of the LDAMP network can be accurately predicted within a short time based on an analytical framework that originates from the state evolution analysis of the AMP algorithm.

\subsection{Data-Aided Model-Driven DL for Channel Feedback}
 In frequency-division duplex massive MIMO systems, an excessive CSI feedback overhead results from a large number of antennas, thereby presenting a significant challenge. To reduce the feedback overhead while maintaining the CSI resolution, many model-driven-based limited feedback schemes have been proposed. However, the conventional model-driven methods face several challenges, including the inaccuracy of the channel model and the inevitable increase in feedback overhead, since the number of antennas is considerably large in massive MIMO systems.

As shown in Fig.\,4, a DL-based network, called CsiNet, has been proposed in \cite{Wen2017Deep} to reduce the feedback overhead in massive MIMO systems. The network architecture of CsiNet is obtained by mimicking the CS architecture, which can be considered as a special case of model-driven DL. CsiNet mainly comprises a convolutional neural network (CNN) that succeeds in image processing and adopts an autoencoder architecture that comprises an encoder for compressive sensing and a decoder for reconstruction. Each RefineNet unit follows the idea of the residual network, that is, it transmits the output of the shallower layer to the input of the deeper layer to avoid gradient vanishing problems in DNNs.

The CsiNet-LSTM in \cite{WangDeep2018} further exploits the time correlation in time-varying massive MIMO channels and to reduce the overhead in the time domain. LSTM is a recurrent neural network (RNN) with memory and expertise in sequence processing and extracting the correlation between adjacent frames in the DL field. Inspired by this approach, the CSI matrices are fed back within the coherence time, share a similar correlation property, and are grouped to create a sequence. All CSI matrices are compressed under a low compression ratio, except for the first CSI matrix of the sequence that is compressed under a higher compression ratio (CR) and reconstructed by CsiNet with a high resolution. The compression is performed because the remaining matrices contain limited information given their similarities to the first matrix. In this way, the feedback overhead can be further reduced. The experiment results show that CsiNet-LSTM is robust to CR reduction and can recover CSI under a low CR with a slight decrease in resolution.

\section{Open Issues}\label{Open}
The model-driven DL for physical layer communications preserves some advantages of the model-driven approaches while greatly reduces the pressure of  accurate modelling. It also retains the powerful learning capability of the DL  and overcomes the requirements for a large amount of training time. Although previous studies have obtained promising results, the model-driven DL for physical layer communications is still in its infancy. In the following, we will discuss several open issues for future investigation, including theoretical analysis, online training, effect of model accuracy, and specialized architecture for model-driven DL.

\subsection{Theoretical Aspects for Model-Driven DL}
The recently developed DL-based communication algorithms have demonstrated a competitive performance but lack unified theoretical foundation and framework, thereby preventing their widespread usage. Nevertheless, performing a theoretical analysis for model-driven DL seems feasible because the algorithms are based on some specific models that always produce rigorous analytical results within certain performance limits. As one of its prominent features, the model-driven DL reduces to an model-driven approach when the elements of DL are removed. Therefore, analyzing the network performance for the model-driven approach is promising. In \cite{CE}, an analytical framework has been constructed based on the performance of the LDAMP-based channel estimation network, which comprises a series of state evolution equations that predict the performance of the network over each layer. Instead of performing a time-consuming Monte Carlo simulation, the behavior can be accurately and quickly predicted and the system design can be optimized by using this analytical framework. Accordingly, a theoretical analysis for model-driven DL needs to be performed in future research.

\subsection{Online Training for Model-Driven DL}
The aforementioned model-driven DL network is trained offline and employed online as most of DL approaches in physical layer communications. The major disadvantage of this approach is that the system remains static because the weights of the neural network do not change when this approach is adopted. Therefore, during the training and design, all possible effects of the future system must be considered, which is very hard in many practical applications. To address this challenge, an online training must be conducted to handle all possible alterations in the environment, which updates its weights during running time \cite{Online_lable}.

As illustrated in Fig.\,5, the model-driven-DL-based OFDM receiver with online and offline training components is considered and both LMMSE channel estimation and zero-forcing detection methods are used as initialized algorithms. The offline-trained network contains two sub-networks trained for indoor and outdoor channels, respectively, in which a large number of learnable parameters are optimized by the stochastic gradient descent algorithm. Parameters $({\boldsymbol \alpha}, {\boldsymbol \beta})$ in Fig.\,5 are additional trainable parameters that are updated via online training to adapt to the fluctuations in the channel conditions. Furthermore, the required labeled data in the online training phase can be recovered from the error-correcting code, which provides perfect label knowledge and avoids the wastage of communication resources.

Fig.\,6 demonstrates the positive effect of the online training for an OFDM  receiver over an outdoor channel.
We consider the influence of different CE network for BER performance and the SD network is same as Fig.\,3. Specifically, the offline-trained CE network in Fig.\,3, the LMMSE method and offline-trained and online-updated network in Fig.\,5 are compared. The performance of the network trained offline for the indoor channel is deteriorated when employed in the outdoor channel, even worse than traditional LMMSE CE method. Apparently, the offline-trained and online-updated network in Fig.\,5 can obtain a huge gain via rapid online training for the outdoor channel because it has two networks trained for indoor and outdoor channel, respectively. The contribution of the two networks is determined by the online training  to adapt to channel alterations when the network is employed.

 \subsection{Effect of Model Accuracy}
In Section \ref{Model}, we illustrate that the model-driven DL is composed of the model, approach (algorithm) and network, and the model only provides a very rough and broad background. However, the degree of model accuracy still influences the performance of model-driven DL approach though DL can compensate the inaccuracy by incorporating side information from the data. A prominent characteristic of the model-driven approach is that when the model is accurate, the optimal solution can be obtained. However, when the model is particularly inaccurate or totally wrong, the performance of the model-driven DL approach  will seriously deteriorate. Therefore, the effect of model accuracy and the ability of DL for compensating the inaccuracy should be investigated in the future.

\subsection{Specialized Model-Driven DL Architecture}
Signals in communication systems are usually processed block-by-block. The model-driven DL architecture used for physical layer communications has an important influence on the performance of each module. However, a specialized network architecture for the communication module is yet to be developed. Most of the existing model-driven DL networks are based on the plain DL architecture or iterative algorithms. However, the CNN and RNN architectures are designed for image and speech signals, respectively, and not specially for physical layer communications. In this case, designing specialized model-driven DL architectures for communication modules becomes desirable.
The LDAMP network proposed in \cite{CE} can  be interpreted as a specialized model-driven DL architecture for channel estimation that inherits the superiority of iterative signal recovery algorithms and uses CNN to learn the channel structure. Therefore, a new and advanced specialized model-driven DL architecture for specific modules in physical layer communications should be investigated.

\section{Conclusion}\label{con}
This article presents a comprehensive overview on model-driven DL in addressing the challenges in physical layer communications. Several recent achievements in transceiver design are highlighted, including transmission schemes, receiver design, and CSI estimation and feedback. Given its impressive capacity and interpreted structure, the model-driven DL shows a competitive performance but with a lower number of trainable variables than the black box architecture and demonstrates potential in intelligent communications.

\section*{Acknowledgements}
This work was supported in part by the National Science Foundation (NSFC) for Distinguished Young Scholars of China with Grant 61625106, and in part by the National Natural Science Foundation of China under Grant 61831013. The work of C.-K. Wen was supported by the Ministry of Science and Technology of Taiwan under Grants MOST 107-2221-E-110-026 and the ITRI in Hsinchu, Taiwan.

\section*{Biographies}

Hengtao He (hehengtao@seu.edu.cn) received the B.S. degree in communications
engineering from Nanjing University of Science and Technology, Nanjing, China, in 2015. He is
currently working towards the Ph.D. degree in information and communications
engineering, Southeast University, China, under the supervision of Prof. Shi Jin. His areas of interests currently include millimeter wave communications, massive MIMO, and machine learning for wireless communications.

Shi Jin [SM'17] (jinshi@seu.edu.cn) received the Ph.D. degree in communications and information systems from Southeast University, Nanjing, in 2007. From June 2007 to October 2009, he was a Research Fellow with the Adastral Park Research Campus, University College London, London, U.K. He is currently with the faculty of the National Mobile Communications Research Laboratory, Southeast University. His research interests include space-time wireless communications, random matrix theory, and information theory. Dr. Jin and his coauthors received the 2010 Young Author Best Paper Award by the IEEE Signal Processing Society and the 2011 IEEE Communications Society Stephen O. Rice Prize Paper Award in the field of communication theory.

Chao-Kai Wen (chaokai.wen@mail.nsysu.edu.tw) received the Ph.D. degree from the Institute of Communications Engineering, National Tsing Hua University, Taiwan, in 2004. He was with Industrial Technology Research Institute, Hsinchu, Taiwan and MediaTek Inc., Hsinchu, Taiwan, from 2004 to 2009. Since 2009, he has been with National Sun Yat-sen University, Taiwan, where he is Professor of the Institute of Communications Engineering. His research interests center around the optimization in wireless multimedia networks.

Feifei Gao [M'09-SM'14] (feifeigao@ieee.org) received the Ph.D. degree from National University of Singapore, Singapore in 2007. He was a Research Fellow with the Institute for Infocomm Research (I2R), A*STAR, Singapore in 2008 and an Assistant Professor with the School of Engineering and Science, Jacobs University, Bremen, Germany from 2009 to 2010. In 2011, he joined the Department of Automation, Tsinghua University, Beijing, China, where he is currently an Associate Professor. Prof. Gao's research areas include communication theory, signal processing for communications, array signal processing, and convex optimizations.

Geoffrey Ye Li (liye@ece.gatech.edu) is a Professor with Georgia Tech. His general research is in signal processing and machine learning for wireless communications. In these areas, he has published over 500 articles with over 34,000 citations and been listed as a Highly-Cited Researcher by Thomson Reuters. He has been an IEEE Fellow since 2006. He won IEEE ComSoc Stephen O. Rice Prize Paper Award and Award for Advances in Communication, IEEE VTS James Evans Avant Garde Award and Jack Neubauer Memorial Award, IEEE SPS Donald G. Fink Overview Paper Award, and Distinguished ECE Faculty Achievement Award from Georgia Tech.

Zongben Xu (zbxu@mail.xjtu.edu.cn)  received his Ph.D. degrees in mathematics from Xi’an Jiaotong University, China, in 1987. He now serves as the Chief Scientist of National Basic Research Program of China (973 Project), and Director of the Institute for Information and System Sciences of the university. He served as Vice President of Xi’an Jiaotong University from 2003 to 2014. He is the owner of Tan Kan Kee Science Award in Science Technology in 2018, the National Natural Science Award of China in 2007, the National Award on Scientific and Technological Advances of China in 2011, CSIAM Su Buchin Applied Mathematics Prize in 2008 and ITIQAM Richard Price Award. He delivered a 45 minute talk on the International Congress of Mathematicians 2010. He was elected as member of Chinese Academy of Science in 2011. His current research interests include intelligent information processing and applied mathematics.

\clearpage

\clearpage

\begin{figure}[!t]
	\centering
	\includegraphics[width=6in]{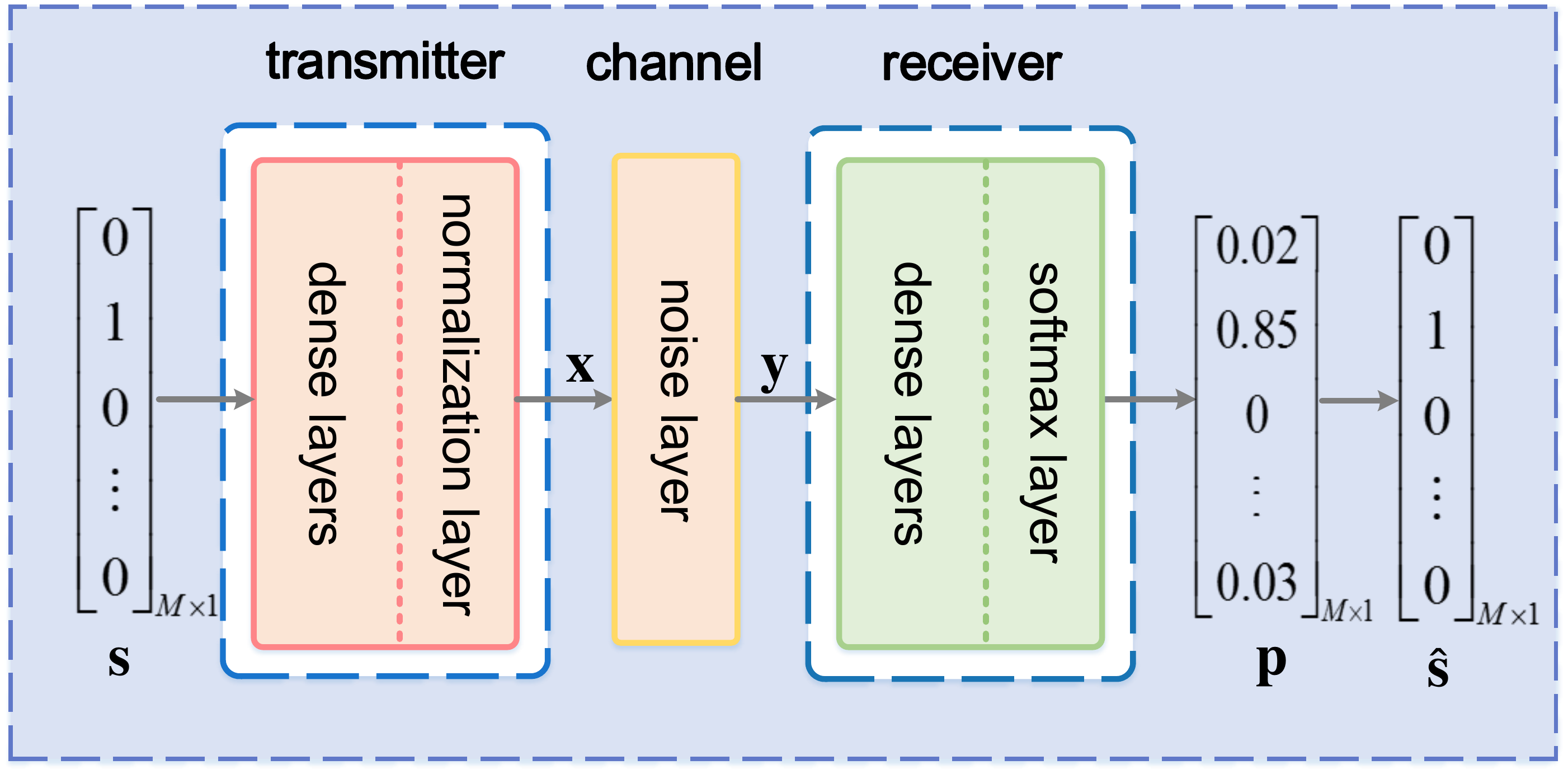}
	\caption{.~~A communication system over an AWGN channel represented as an autoencoder in \cite{OShea}. The $\mathbf{s}$ in one-hot representation is encoded to an N-dimensional transmitted signal $\mathbf{x}$. After adding channel noise, the noisy encoded signal $\mathbf{y}$ is then decoded to an M-dimensional probability vector $\mathbf{p}$ before determining $\mathbf{s}$.}
	\label{autoencoder}
\end{figure}
\clearpage

\begin{figure*}[!t]
	\centering
	\includegraphics[width=6in]{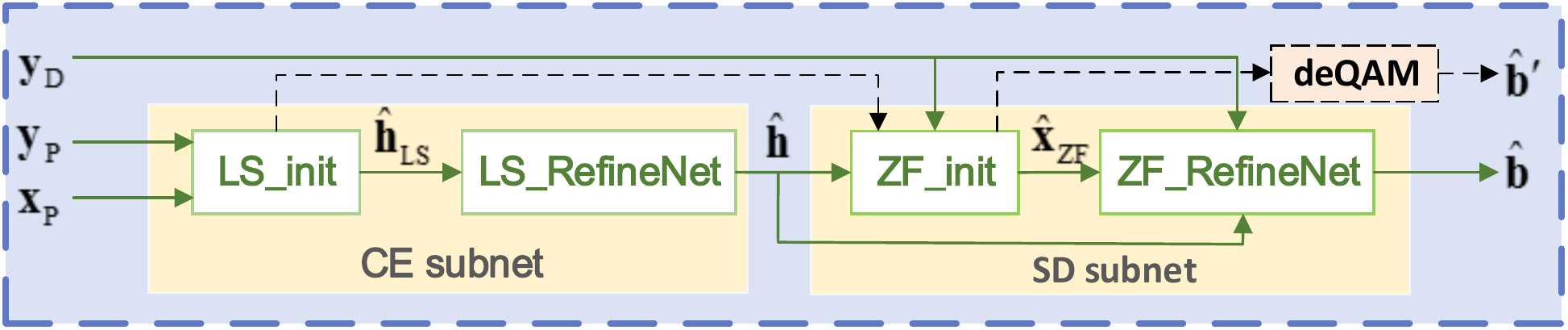}
	\caption{.~~ComNet receiver architecture. The two subnets use traditional communication solutions as initializations and apply DL networks to refine the coarse inputs. The dotted shortpath provides a relatively robust candidate for the recovery of binary symbols.}
	\label{ComNet}
\end{figure*}
\clearpage

\begin{figure*}[!t]
		\centering
		\includegraphics[width=2in,height=3in]{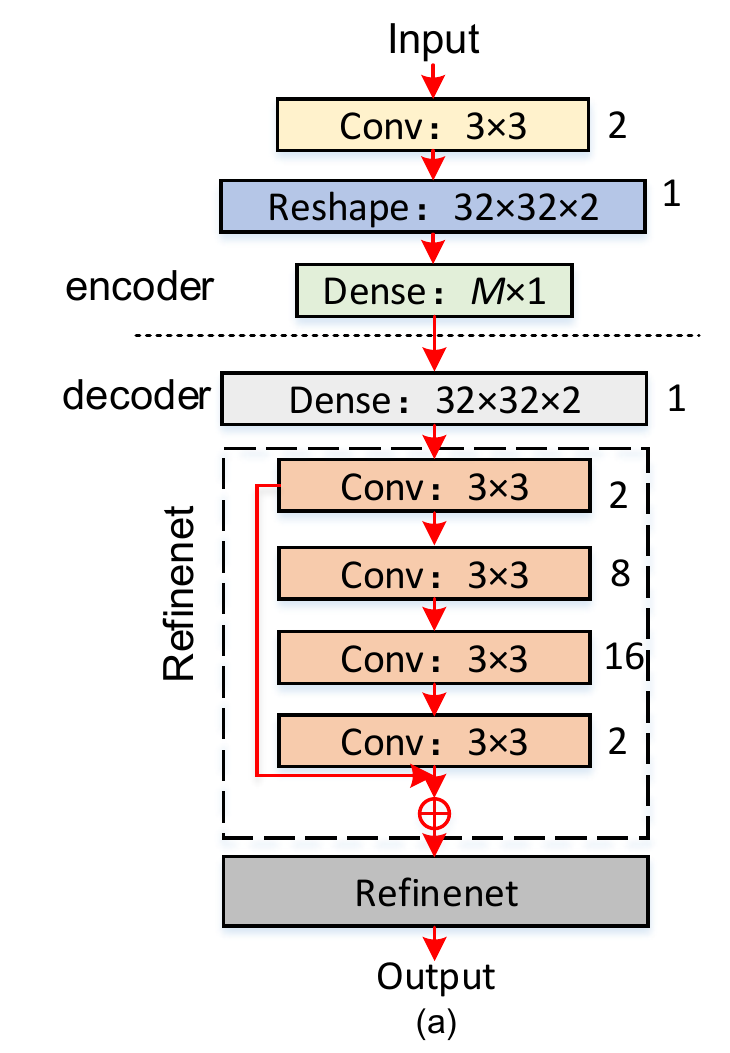}\includegraphics[width=4in,height=3in]{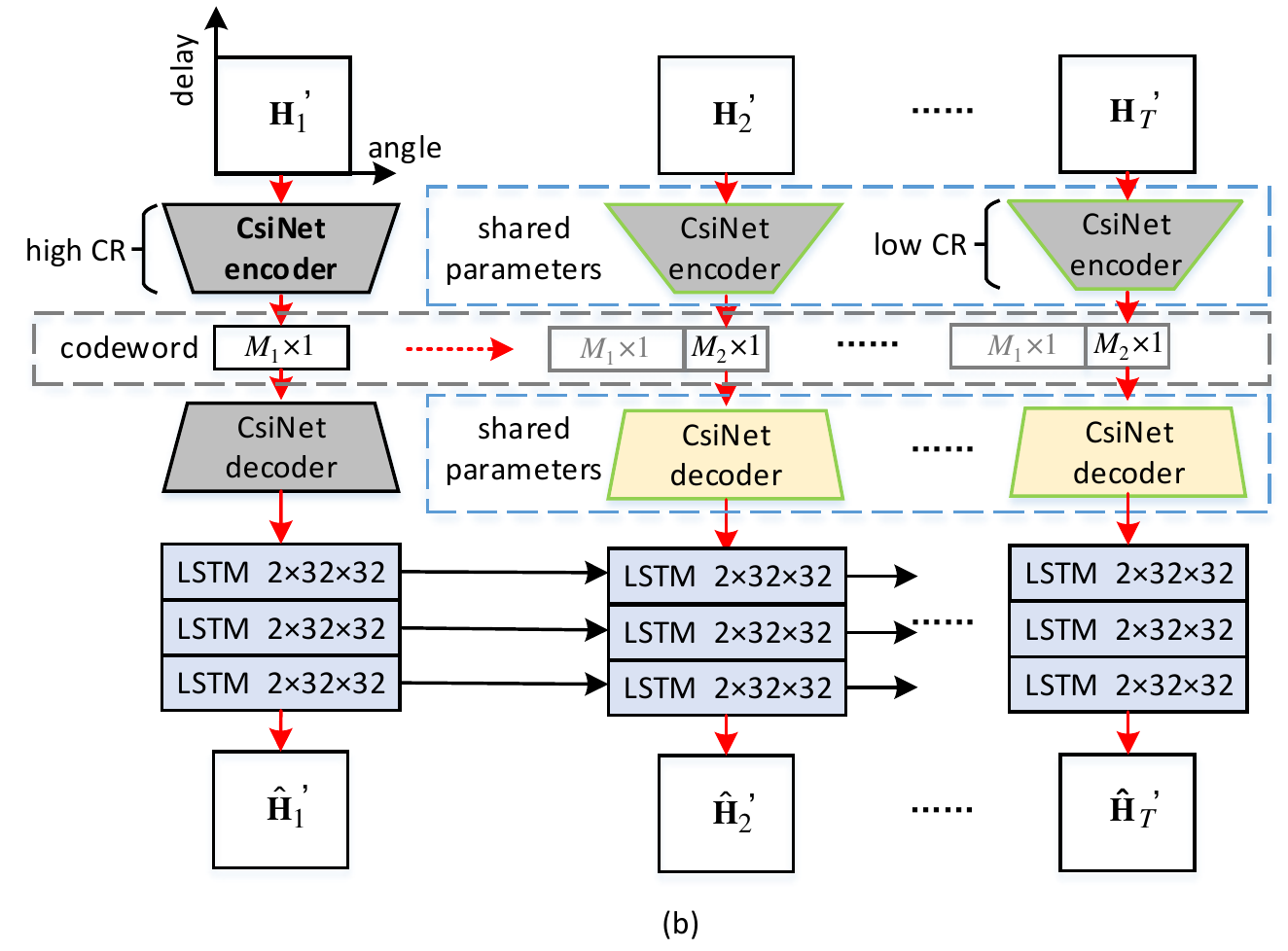}
		\vspace{-0.3cm}
		\caption{(a) CsiNet architecture in \cite{Wen2017Deep}. This architecture comprises an encoder with a $ 3 \times 3 $ conv layer and an $M$-unit dense layer for sensing as well as  a decoder with a $ 2N_{c}'N_{t} $-unit dense layer and two RefineNet for reconstruction.  Each RefineNet contains four $ 3 \times 3 $ conv layers with different channel sizes. (b) Overall architecture of CsiNet-LSTM in \cite{WangDeep2018}. ${\bf H}_{1}'$ and the remaining ${T-1}$ channel matrices are compressed by high-CR and low-CR CsiNet encoders, respectively. Code words are concatenated before fed into the low-CR CsiNet decoder, and the final reconstruction is performed by three $ 2N_{c}'N_{t} $-unit LSTMs. }		
				\vspace{-0.5cm}
		\label{system architecture}
\end{figure*}
\clearpage

\begin{figure*}[!t]
	\centering
	\includegraphics[width=6in]{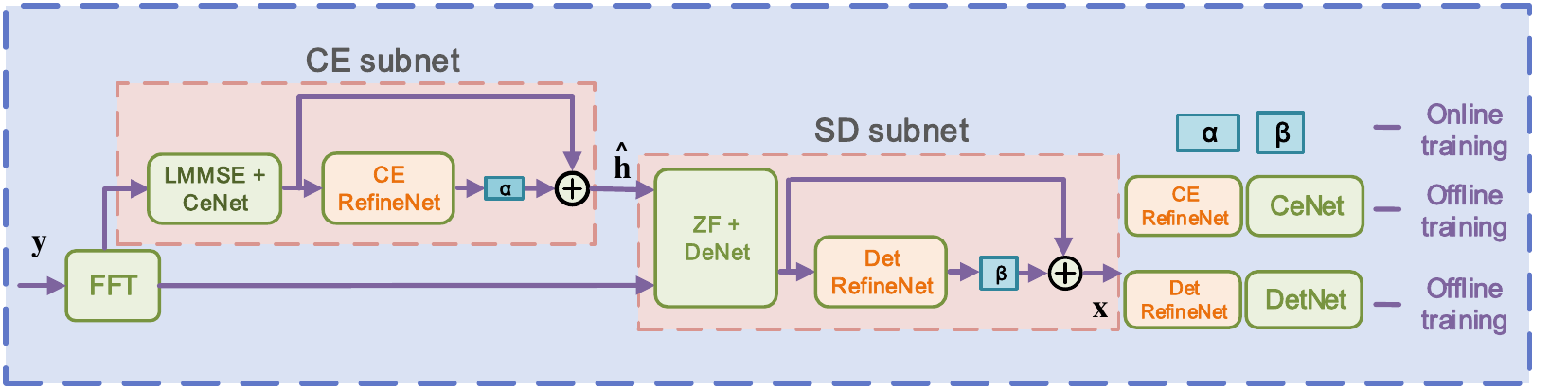}
	\caption{.~~OFDM receivers including online and offline training.}
	\label{Online}
\end{figure*}
\clearpage

\begin{figure*}[!t]
	\centering
	\includegraphics[width=6in]{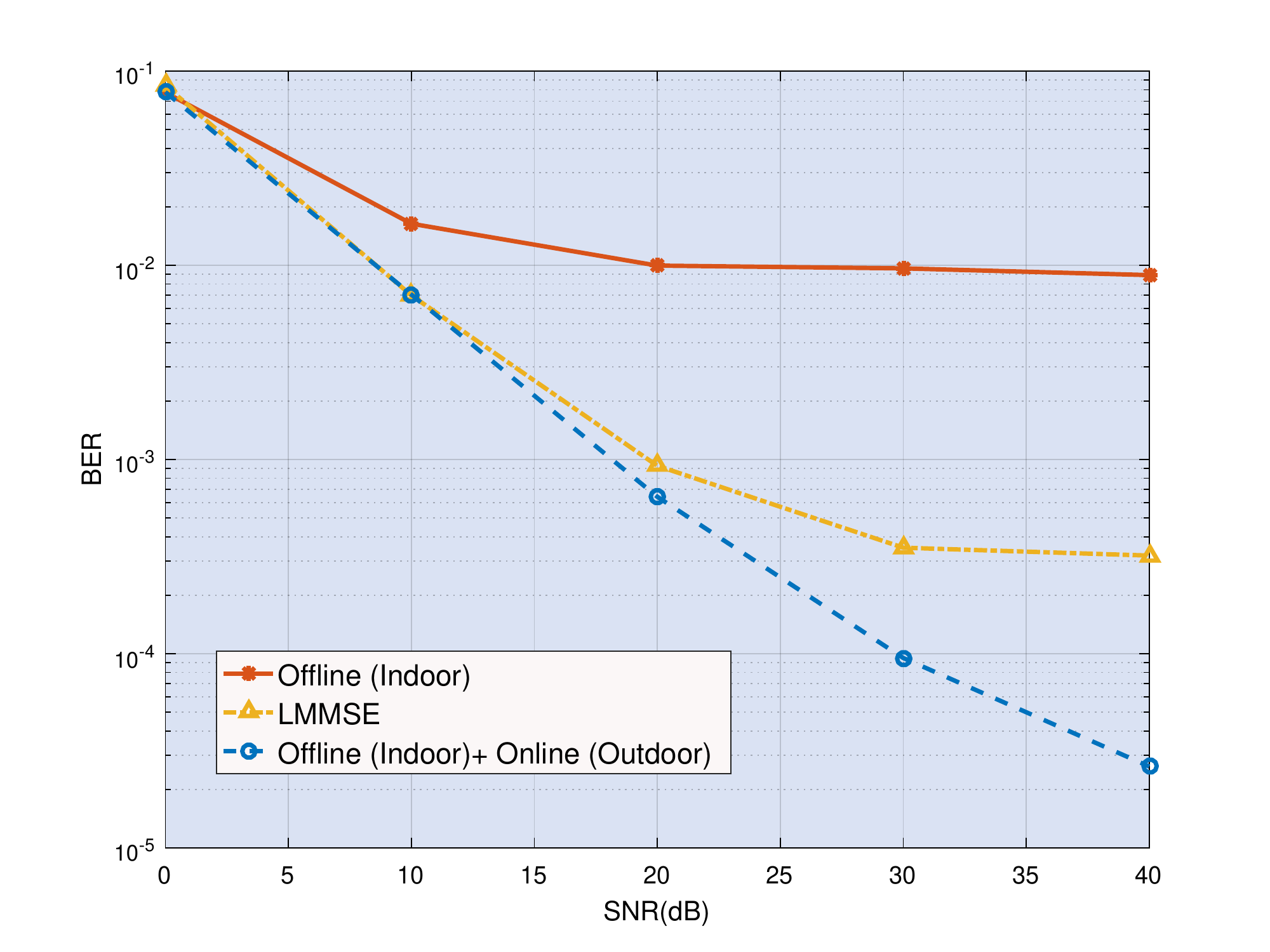}
	\caption{.~~BER Performances of OFDM receivers with online and offline training.}
	\label{Online}
\end{figure*}

\end{document}